\documentclass{article}

\usepackage{arxiv}

\usepackage[utf8]{inputenc} 
\usepackage[T1]{fontenc}    
\usepackage{hyperref}       
\usepackage{url}            
\usepackage{booktabs}       
\usepackage{amsfonts}       
\usepackage{nicefrac}       
\usepackage{microtype}      
\usepackage{lipsum}		
\usepackage{graphicx}
\usepackage[english]{babel}
\usepackage{doi}

\usepackage{algorithm}
\usepackage{amsmath} 
\usepackage[capitalize]{cleveref} 
\newcommand{\norm}[1]{\left\lVert#1\right\rVert}
\newcommand{\lword}[1]{\leavevmode\nobreak\hskip0pt plus\linewidth\penalty50\hskip0pt plus-\linewidth\nobreak#1}

\title{Benchmarking quantum machine learning kernel training for classification tasks}

\author{ \href{https://orcid.org/0000-0001-5790-0577}{\includegraphics[scale=0.06]{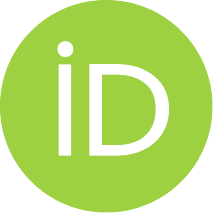}\hspace{1mm}Diego Alvarez-Estevez} \\
	CITIC Research Center \\
	Universidade da Coruña \\
	A Coruña, 15071 \\
	\texttt{diego.alvareze@udc.es} \\
}

\date{}

\renewcommand{\headeright}{}
\renewcommand{\undertitle}{}
\renewcommand{\shorttitle}{QKT benchmarking}

\begin{document}
\maketitle

\begin{abstract}
Quantum-enhanced machine learning is a rapidly evolving field that aims to leverage the unique properties of quantum mechanics to enhance classical machine learning. However, the practical applicability of these methods remains an open question, particularly beyond the context of specifically-crafted toy problems, and given the current limitations of quantum hardware.
This study focuses on quantum kernel methods in the context of classification tasks. In particular, it  examines the performance of Quantum Kernel Estimation (QKE) and Quantum Kernel Training (QKT) in connection with two quantum feature mappings, namely ZZFeatureMap and CovariantFeatureMap. Remarkably, these feature maps have been proposed in the literature under the conjecture of possible near-term quantum advantage and have shown promising performance in ad-hoc datasets. In this study, we aim to evaluate their versatility and generalization capabilities in a more general benchmark, encompassing both artificial and established reference datasets. Classical machine learning methods, specifically Support Vector Machines (SVMs) and logistic regression, are also incorporated as baseline comparisons.
Experimental results indicate that quantum methods exhibit varying performance across different datasets. Despite outperforming classical methods in ad-hoc datasets, mixed results are obtained for the general case among standard classical benchmarks. Our experiments call into question a general added value of applying QKT optimization, for which the additional computational cost does not necessarily translate into improved classification performance. Instead, it is suggested that a careful choice of the quantum feature map in connection with proper hyperparameterization may prove more effective.
\end{abstract}


\section{Introduction}
In the crossover of Artificial Intelligence and Quantum Computing, the emerging field of Quantum Machine Learning (QML) arises with the aim to harness the distinctive capabilities of quantum mechanics to boost capabilities of current state-of-the-art (classical) machine learning methods. Work in this field is rapidly evolving, and considerable progress has been made in the recent years. However, to what extend QML might live to its expectations is yet a subject of debate. This motivates the necessity of carrying out benchmark studies that provide a structured framework to assess the performance and potential use of these methods.
Indeed, there are still many open questions and challenges to be addressed, in particular, in relation to the limited capabilities of real quantum computing hardware available today. In this context, benchmarking models via classical simulations on small, but heterogeneous, and still challenging datasets might offer valuable insights to drive the field forward before more capable quantum hardware becomes available. 
Following this line, this work carries out experimentation in the context of quantum kernel methods to test the performance of the so-called \textit{quantum kernel alignment} technique \cite{hubregtsen_training_2022, glick_covariant_2024}. For that purpose, benchmarking is performed that involves different artificial and small-scale reference classification datasets to compare both quantum realizations with their classical counterparts. 

The specific interest in kernel methods arises in part as it has recently been shown that variational supervised QML models (often referred to as "quantum neural networks") fundamentally reduce to kernel methods. However, kernel methods come with additional advantages, such as the guarantee of finding better or equally good solutions, convex optimization, or even improved performance under certain conditions \cite{schuld2021_are_kernel_methods}. Moreover, quantum kernel methods can lead to a provable separation between classical and quantum learners for certain problems, opening the door for possible quantum advantage achievable even in the current Noisy Intermediate-Scale Quantum (NISQ) era \cite{liu_rigorous_2020}.

Motivated by these properties, the literature on potential practical applications of quantum kernel methods is growing rapidly. Specific use cases range from finance \cite{miyabe2023quantummultiplekernellearning}, fraud detection \cite{Grossi2022_fraud_detection, DiPierro2021}, healthcare \cite{Krunic_2022}, earth sciences \cite{Miroszewski_10214376, Wijata_10641442, Miroszewski_10282416}, or physics \cite{Wu_2022MG, Peters2021MachineLO}, among others \cite{Huang_2021, Mengoni2019KernelMI}. However, there is a need for benchmark studies that systematically assess the performance analysis of these methods from a more general perspective \cite{DiMarcantonio2022QuantumAS, bowles_better_2024}.

Following this line, this study contributes to the existing body of work on performance evaluation by following a general benchmarking approach that involves alternative datasets and methods. In particular, we take as reference two quantum feature maps, namely ZZFeatureMap and CovariantFeatureMap, for which the potential for quantum advantage has been postulated in the context of quantum kernel estimation using ad-hoc data \cite{havlicek_supervised_2019, glick_covariant_2024}. However, a more extensive evaluation of these methods' capabilities on general-purpose benchmarks, including a comparison to baseline classical counterparts, is still necessary. This is particularly true in the case of CovariantFeatureMap. 
We methodically examine the impact of quantum (alignment) kernel training on the generalization performance of the resulting models. We follow two distinct parameterization strategies to this end: "shared" and "dedicated", applicable to both feature mappings. To the best of our knowledge, the impact of these strategies on model generalization has not been previously evaluated in the literature. Furthermore, we employ the "weighted" version of target alignment, in contrast to the prevailing practices that utilize the "unweighted" alignment version derived from the seminal work of Hubregsten et al. \cite{hubregtsen_training_2022}. As discussed in Glick et al. \cite{glick_covariant_2024}, this "weighted" version theoretically optimizes the SVM upper bound on the generalization error. However, the benchmarking of weighted kernel alignment on a wide set of datasets has not yet been performed.

\section{Methods}
\label{sec:methods}
We follow with a concise exposition of kernel methods from a general perspective, with particular emphasis on \textit{Support Vector Machines} in the context of classification problems. We then describe the use of quantum computers for the estimation of kernel entries, a process referred to as \textit{Quantum Kernel Estimation}. Finally, we show how we can extend this approach with parameterized adaptation, resorting to \textit{Quantum Kernel Training}.

\subsection{Kernel methods}
\label{subsec:kernel_methods}
Kernel methods are an important class of algorithms in machine learning. The underlying idea is that of transforming the input data into a higher-dimensional space where the problem at hand becomes easier to solve. This transformation (i.e., \textit{feature mapping}) is done using a \textit{kernel function}, which computes the inner product of two inputs as a measure of their similarity in the induced higher-dimensional space. This procedure is core for a wide range of methods that can be expressed in terms of these operations. The beauty of kernel methods is that the kernel function is designed in such a way that it implicitly computes the inner-product in the high-dimensional space, however without ever explicitly accessing the coordinates of the data in that space, which often is computationally expensive. This is known as the "kernel trick", and for this to be possible the mapping function has to meet the so-called \textit{Mercer's conditions} \cite{bishop2007}. 

Kernel methods find application in various areas and cover many machine learning settings, including both supervised and unsupervised learning. They are particularly useful when data are not linearly separable in their original space. By transforming data into a higher-dimensional space, kernel methods allow linear algorithms to be applicable to solve non-linear problems. Here we focus in which arguably is the most famous algorithm within kernel methods, the Support Vector Machine (SVM). More specifically, with application to classification problems, leading to the so-called Support Vector Classifier (SVC).

Let us consider a binary classification dataset of samples $\{\overrightarrow{x_{i}}\}_{i=1...m}$ from input space $\Omega\subset\mathbb{R}^{d}$, with corresponding class labels $y_{i} \in \{-1,1\}$. The binary SVC classifies examples according to

\begin{equation} 
\label{eq:1}
y(\overrightarrow{x})=sgn(\langle\overrightarrow{w},\phi(\overrightarrow{x})\rangle+b)
\end{equation}

where the vector $\overrightarrow{w}\in\mathbb{R}^{d_{f}}$ represents the hyperplane normal to the linear decision boundary that separates the two classes, and lives in the induced feature space that results from the mapping $\phi$ (and thus, usually $d_{f}>>d$) and $b\in\mathbb{R}$ is the intercept, also referred as \textit{offset} or \textit{bias} term.

Training the classifier following the SVM formulation corresponds to finding $\overrightarrow{w}$ perpendicular to the separating hyperplane with maximum possible distance between samples lying in the margins of the two classes (also called \textit{support vectors}). Mathematically, this equals to solving the following quadratic problem:

\begin{equation} 
\label{eq:2}
\begin{split}
\text{minimize} & \quad\frac{1}{2}\norm{\overrightarrow{w}}^{2}_{2}+C\sum_{i=1}^{m}\xi_{i} \\
\text{subject to:} & \quad y_{i}(\langle\overrightarrow{w},\phi(\overrightarrow{x})\rangle+b\geq1-\xi_{i} \\
 & \quad \xi_{i}\geq0 \quad \forall i=1,\ldots, m
\end{split}
\end{equation}

with $C\geq0$ acting as regularization parameter for datapoints that are not linearly separable (even in the feature space), for which additional slack variables $\xi_{i}$ are introduced acting as penalization terms. This is a convex optimization problem, for which by applying Lagrangian formulation, and noticing that we can express $\overrightarrow{w}$ as weighted sum of the embedded data points $\overrightarrow{w}=\sum_{i}^{m}\alpha_{i}\phi(\overrightarrow{x_{i}})$, the following equivalent Wolfe dual version is obtained

\begin{equation} 
\label{eq:3}
\begin{split}
\text{maximize} & \quad\sum_{i=1}^{m}\alpha_{i}-\frac{1}{2}\sum_{i,j=1}^{m}\alpha_{i}\alpha_{j}y_{i}y_{j}\langle\phi(\overrightarrow{x_{i}}),\phi(\overrightarrow{x_{j}})\rangle \\
= & \quad\sum_{i=1}^{m}\alpha_{i}-\frac{1}{2}\sum_{i,j=1}^{m}\alpha_{i}\alpha_{j}y_{i}y_{j}\kappa(\overrightarrow{x_{i}},\overrightarrow{x_{j}}) \\
\text{subject to:} & \quad \sum_{i=1}^{m}y_{i}\alpha_{i}=0
 \\
 & \quad 0\leq\alpha_{i}\leq\text{C} \quad \forall i=1...m
\end{split}
\end{equation}

in which application of the "kernel trick" results straightforward, and noticing that for any new data point $\overrightarrow{z}$ not included in the training set, the classification decision can be efficiently calculated as

\begin{equation} 
\label{eq:4}
y(\overrightarrow{z})=sgn(\sum_{i=1}^{m}y_{i}\alpha_{i}\kappa(\overrightarrow{x_{i}},\overrightarrow{z})+b)
\end{equation}

\subsection{Quantum Kernel Estimation}

Variational Quantum Algorithms, a hybrid quantum-classical approach that relies on the use of Parameterized Quantum Circuits (PQCs), represent the current workhorse of quantum machine learning methods in the current NISQ era. 

Quantum Kernel Estimation (QKE) is a subclass of these methods where the kernel function is calculated using a PQC that acts as \textit{quantum feature map}, embedding datapoints in the original space (classical data) into the induced Hilbert space of quantum states \cite{Schuld2019_QML_in_Hilbert, havlicek_supervised_2019}. Let us denote by $U(\vec{x})$ the unitary representation of the PQC that acts on \textit{n}-qubits initialized to the ket-zero state, the following data-dependent state vector results: 

\begin{equation} 
\label{eq:5}
|\phi(\vec{x})\rangle=U(\vec{x})|0^{\otimes n}\rangle
\end{equation}

The quantum state vector representation, however, does only have physical meaning up to global phase. This is problematic, for which the inner product of two physically identical state vectors might result in different values. Thus, in practice, the mapping is considered in the space of density matrices, which for pure states is given by:

\begin{equation}
\label{eq:6}
\Phi:\vec{x} \in \Omega \rightarrow \rho(\vec{x})=|\phi(\vec{x})\rangle\langle\phi(\vec{x})|
\end{equation}

Hence, given two samples $\vec{x_{i}}, \vec{x_{j}}$ and using the Hilbert-Schmidt inner product between the corresponding density matrices $\rho(\vec{x_{i}})$ and $\rho(\vec{x_{j}})$, $\langle\rho(\vec{x_{i}}),\rho(\vec{x_{j}})\rangle=Tr\{\rho(\vec{x_{i}})\rho(\vec{x_{j}})\}$ the following kernel function naturally derives:

\begin{equation}
\label{eq:7}
\kappa(\vec{x_{i}},\vec{x_{j}})=|\langle\phi(\vec{x_{j}})|\phi(\vec{x_{i}})\rangle|^2
\end{equation}

which corresponds to the \textit{fidelity} or \textit{overlap} between the corresponding $|\phi(\vec{x_{i}})\rangle$ and $|\phi(\vec{x_{j}})\rangle$ quantum states. Following \eqref{eq:5}:

\begin{equation}
\label{eq:8}
\kappa(\vec{x_{i}},\vec{x_{j}})=|\langle0^{\otimes n}|U\textsuperscript{\textdagger}(\vec{x_{j}})|U(\vec{x_{i}})|0^{\otimes n}\rangle|^2
\end{equation}

There are several algorithms to estimate the overlap of two arbitrary quantum states. One straightforward approach is to use a \textit{SWAP} circuit. It has the advantage that works with mixed states, but we need to use some auxiliary qubits that result in a circuit of width $2n+1$. Also, depending on the quantum hardware architecture, extra depth might be required for implementation of the required controlled SWAP operations. However, because in this case we know exactly how the state was created, an alternative approach is to apply the encoding circuit, followed by the adjoint of its unitary, $U\textsuperscript{\textdagger}(\vec{x})$, and finally perform measurement to record the frequency the output is found to be $|0^{\otimes n}\rangle$. That is, the approach literally follows optimization expression \eqref{eq:3}, and results in a circuit of depth $2n$. However, one has to bear in mind that this protocol is only applicable in the context of pure states \cite{havlicek_supervised_2019}. 

Among the nice properties of QKE we have that as the kernel computation does only represent a small subroutine in the general hybrid pipeline, usually long coherence times are not required. In addition, because kernel values can be estimated using circuits short-depth circuits, it makes of it an ideal candidate in the context of current NISQ devices. Furthermore, regularization and error-mitigation techniques have been discussed that make QKE estimation method described above robust even in the presence of certain amount of noise \cite{havlicek_supervised_2019,hubregtsen_training_2022}.

One crucial consideration in the pursuit of potential quantum advantage, in terms of infeasibility of classical simulation, regards the complexity of the quantum circuit. In this regard, it is important that the underlying represented state does not simplify to a \textit{product state}. QKE may also offer additional benefits beyond the discussion on the achievement of possible quantum speed-ups. These include the potential to access more expressive and possibly better separable induced feature spaces. This subject is currently an active area of research.

\subsection{Quantum Kernel Training}
\label{subsec:QKT}

One should notice that under the QKE technique described in the previous section, the unitary $U(\vec{x})$ remains purely dependent on input data, and it admits no further optimization once the specific structure of the representative parameterized quantum circuit has been chosen. However, the proper choice of the optimal feature map depends on the specific targeted problem in practice. To partially mitigate this challenge, one might consider to fit additional parameters $\theta$ for a variational family of kernels, leading to Quantum Kernel Training (QKT) technique that is introduced next.

\subsubsection{Classical background}

Classical kernel-target alignment technique relies on the idea of using of a proxy quantity, namely \textit{alignment}, that enables efficient evaluation of the appropriateness of choosing a specific kernel function for the targeted learning problem \cite{NIPS2001_1f71e393}. Indeed, the success of kernel methods is very much dependent on the appropriate choice of a kernel function such that the geometrical structure of the mapped data in the feature space fits the specific learning target. In the case of SVMs, that would conceptually translate to the possibility of achieving maximum margin linear separability between the target classes in the induced feature space. 

Let $K_1$ and $K_2$ be two kernel matrices, the empirical alignment derives from evaluating and normalizing their inner product, that is 

\begin{equation}
\label{eq:9}
A(K_1,K_2)=\frac{\langle\text{K}_1,\text{K}_2\rangle}{\sqrt{\langle\text{K}_1,\text{K}_1\rangle\langle\text{K}_2,\text{K}_2\rangle}}
\end{equation}

which corresponds to calculation of to their \textit{cosine similarity} in the case $\langle\text{K}_1,\text{K}_2\rangle$ is a real non-negative value, condition that we have guaranteed for which valid kernel matrices are positive semi-definitive.

Given a parameterized kernel family function $\kappa_\theta$, Cristiani et al. \cite{NIPS2001_1f71e393} have shown that, by choosing $K_{1}=\kappa_{\theta}(\vec{x_{i}},\vec{x_{j}})$ and $K_{2}=y_{i},y_{j}$, and optimizing (maximizing) $A(K_1,K_2)$ on the training set, one is expected to keep high alignment and achieve good classification generalization on the test set as well. Of note, substituting in \eqref{eq:9} and operating:

\begin{equation}
\label{eq:10}
TA(\kappa_{\theta})=\frac{\sum_{i,j=1}^{m}y_{i}y_{j}\kappa_{\theta}(\overrightarrow{x_{i}},\overrightarrow{x_{j}})}{\sqrt{(\sum_{i,j=1}^{m}\kappa_{\theta}(\overrightarrow{x_{i}},\overrightarrow{x_{j}})^2)(\sum_{i,j=1}^{m}y_{i}^2y_{j}^2)}}
\end{equation}

Hence, iterative optimization methods can be applied to choose the best parameter set $\vec{\theta_{opt}}$ that maximizes $TA(\kappa_{\theta})$.
This is in contrast to the more complex (often widespread) approach of considering kernel optimization as a problem of model selection using uninformed cross-validation approaches that involve computationally intensive training of kernel machines. For a more comprehensive overview of (classical) kernel alignment, its mathematical justification, and applications, we refer to the nice review by Wang et al. \cite{wang_overview_2015}.
 
\subsubsection{Quantum approximations}
\label{subsubsec_quantum_approximations}

Inspired by ideas on classical kernel-target alignment, QKT has been proposed as hybrid quantum-classical optimization algorithm in the context of QKE \cite{hubregtsen_training_2022}. In this setting kernel-target alignment is used as cost function from which to iteratively derive new parameters for an embedding quantum circuit, which is dependent on both the input data points $\vec{x}$ and variational parameters $\vec{\theta}$. The \textit{quantum processing unit} (QPU) is used at each variational step to obtain the corresponding kernel matrix entries $\kappa_{\theta}(\vec{x_{i}},\vec{x_{j}})$. That is, the data embedding process described in \eqref{eq:5} now becomes:

\begin{equation} 
\label{eq:11}
|\phi_{\theta}(\vec{x})\rangle=U_{\theta}(\vec{x})|0^{\otimes n}\rangle
\end{equation}

and the associated QKE routine:

\begin{equation}
\label{eq:12}
\kappa_{\theta}(\vec{x_{i}},\vec{x_{j}})=|\langle0^{\otimes n}|U_{\theta}\textsuperscript{\textdagger}(\vec{x_{j}})|U_{\theta}(\vec{x_{i}})|0^{\otimes n}\rangle|^2
\end{equation}

Alternatively, the use of \textit{weighted} kernel alignment has been recently introduced for the optimization of the \textit{fiducial state} in the context of (quantum) \textit{covariant quantum kernels} for data spaces that have a group structure \cite{glick_covariant_2024}. 

Under their approach, the SVM optimization problem \eqref{eq:3} directly translates to

\begin{equation} 
\label{eq:13}
\min_{\vec{\theta}}\max_{\vec{\alpha}}\text{F}(\vec{\alpha}, \vec{\theta})=\sum_{i=1}^{m}\alpha_{i}-\frac{1}{2}\sum_{i,j=1}^{m}\alpha_{i}\alpha_{j}y_{i}y_{j}\kappa_{\theta}(\overrightarrow{x_{i}},\overrightarrow{x_{j}})
\end{equation}

subject to the same $\alpha_{i}$ constraints. 

Indeed, noticing from \eqref{eq:10} that 

\begin{equation} 
\label{eq:14}
TA(\kappa_{\theta})\quad\propto\sum_{i,j=1}^{m}y_{i}y_{j}\kappa_{\theta}(\overrightarrow{x_{i}},\overrightarrow{x_{j}})
\end{equation}

it derives that the second term of \eqref{eq:13} involves the kernel alignment 
\textit{weighted} by the Lagrange multipliers, hence the name. Notably, whereas the \textit{unweighted} version of the alignment c.f. \eqref{eq:10} implies finding kernels that maximize the intra-class and minimize the inter-class overlap for all points included in the training set, its \textit{weighted} counterpart aims as well to maximize the margin gap around the decision boundary. Consequently, only the subset of support vectors contribute to the alignment (for all the rest $\alpha_{i}=0$). Also remarkable, because of the negative sign in the second term, solving the quadratic SVM dual problem involves minimizing over the kernel parameters $\theta$. It can be shown that solving this min-max problem corresponds to finding $\theta_{opt}$ such that $\kappa_{\theta_{opt}}$ minimizes the upper bound of the SVM generalization error \cite{glick_covariant_2024}. In general, we also note that kernel optimization using the kernel-target alignment is closely related to the “metric learning” approach proposed in \cite{lloyd2020metriclearning} for the training of quantum feature embeddings.

For solving problem \eqref{eq:13}, we follow a variant of the stochastic classical-quantum iterative algorithm proposed by the same authors. The algorithm evaluates kernel matrices on a quantum processor and updates parameters $\theta$ and $\alpha$ for the next iteration with classical optimizers. The use of Simulataneous Perturbation Stochastic Approximation (SPSA) is proposed for the evolution of $\theta$, for which minimization of \eqref{eq:13} with respect to the kernel parameters leads to highly non-convex landscape in general. For this purpose, the method approximates the gradient of the loss with respect to $\theta$ using only two objective function evaluations at each iteration, regardless of the dimension of $\theta$. This involves evaluating the corresponding kernel matrices, which is done using the QPU following \eqref{eq:12}. Once kernel matrices have been calculated, standard SVM classical solvers can be used for maximization with respect to $\alpha$ in the $min-max$ problem according to \eqref{eq:13}, which remains concave. One additional advantage SPSA regards its robustness in the context of noisy objective functions, which is convenient given kernels are evaluated on noisy quantum hardware. 

\section{Experimental setting}
\label{sec:experiments}

The main focus of our experimentation (and our work) is dual-purpose. First, we would like to test the versatility and generalization capabilities of some specific quantum feature mapping circuits that have been proposed in the context of QML-related literature (and more specifically, QKE), for which the possibility of quantum advantage has been conjectured \cite{havlicek_supervised_2019, glick_covariant_2024}. The quantum feature mappings in question have been shown to be able to achieve perfect separability on specifically targeted (ad-hoc) classification learning problems. However, in the related reference works, no attempt was performed to solve the same task using classical learning counterparts. Moreover, it remains an open question to which extend these mappings could find some practical application in solving other learning tasks beyond the aforementioned artificially-generated datasets. For this purpose, in addition to trying to reproduce results shown in these seminal works, we would like to expand the benchmarking to include additional well-known classical machine learning datasets, as well as the use of classical SVMs as baseline methods. Second, and taking as reference the same benchmark database, we would like to evaluate the contribution of QKT using weighted alignment following the methods described in Section \ref{subsec:QKT}. 

\subsection{Feature mappings}

In this work we are taking as reference two quantum feature maps, namely \textit{ZZFeatureMap} and \textit{CovariantFeatureMap}. 

The first one, related to the \textit{IQP} circuit family introduced in \cite{Bremner2016} (and thus sometimes also regarded as as \textit{IQP}-circuit throughout the literature), derives from a family of encoding circuits that transform input data $\vec{x}\in\mathbb{R}^{d}$ applying the following \textit{n}-qubit unitary to evolve from the $|0\rangle^{\otimes n}$ initialization state \cite{havlicek_supervised_2019}:

$$ U(\vec{x})=\prod_d PE_{\Phi(\vec{x})}H^{\otimes n}$$

with \textit{PE} denoting the Pauli Expansion $$\ PE_{\Phi(\vec{x})}=\exp\left(i\sum_{S\subseteq[n]}\phi_S(\vec{x})\prod_{k\in S} P_i\right)\text{,}$$

where $$\quad\phi_S:\vec{x}\mapsto \Bigg\{
    \begin{array}{ll}
        x_i & \text{if}\ S=\{i\} \\
        (\pi-x_i)(\pi-x_j) & \text{if}\ S=\{i,j\}
    \end{array}$$

and in which $P_i \in \{ I, X, Y, Z \}$, i.e., the set of Pauli matrices, $S \in \{\binom{n}{k}\ \mathrm{combinations,\ }k = 1,... n \}$ describes the qubit connectivity, and $d$ the number of repetitions that determine the circuit depth. More specifically, the \textit{ZZ-FeatureMap} results from choosing $d=2$, $k=2$, $P_0 = Z, P_1 = ZZ$, leading to the following relatively shallow circuit which has been conjectured to be hard to estimate classically \cite{havlicek_supervised_2019}:

\begin{equation}
\label{eq:15}
U_{ZZ}(\vec{x}) = \left[ \exp\left(i\sum_{jk} \phi_{\{j,k\}}(\vec{x}) \, Z_j \otimes Z_k\right) \, \exp\left(i\sum_j \phi_{\{j\}}(\vec{x}) \, Z_j\right) \, H^{\otimes n} \right]^2\text{.}
\end{equation}

The second quantum feature map at test, i.e., \textit{Covariant-FeatureMap}, derives from the already mentioned work on covariant quantum kernels for data with group structure \cite{glick_covariant_2024}. The motivational hypothesis is that advanced classical kernels for group-data are computationally expensive to evaluate, and thus in practice one has to resort to approximation methods for their computation. Thus a potential advantage of quantum kernels is that such approximation may not be necessary \cite{glick_covariant_2024}. This hypothesis is supported by previous work showing that covariant quantum kernels can lead to provable quantum advantage over classical learners for specifically constructed learning problems \cite{liu_rigorous_2020}. Generalizing this result, Glick et al. proposed \textit{covariant feature maps} characterized as $U_{\theta}(\vec{x})=D(\vec{x})V(\vec{\theta})$, with $D(\vec{x})$ corresponding to the unitary representation of a group $G$, with $\vec{x}\in G$, and $V(\vec{\theta})$ representing a parameterized (fiducial) state, such that $|\phi_{\theta}(\vec{x})\rangle=D(\vec{x})|\phi_{\theta}\rangle=D(\vec{x})V(\vec{\theta})|0^{\otimes n}\rangle$. 

More specifically, based on the synthetic \textit{labeling cosets with error} (LCE) problem, the specific parameterized (\textit{ad-hoc}) mapping was characterized by choosing: 

\begin{equation}
\label{eq:16}
    D(\vec{x})=\otimes_{k=1}^{n}R_{X}(x_{2k-1})R_{Z}(x_{2k})\text{,}
\end{equation}

and

\begin{equation}
\label{eq:16_bis}
    \text{V}(\theta)=\prod_{(k,t)\in\text{E}}CZ_{k,t}\prod_{k\in\mathcal{V}}R_{Y_{k}}(\theta)
\end{equation}

where $R_{X}$, $R_{Y}$ stand for the single-qubit rotation operators around the corresponding axes of the Bloch sphere, with $R_{Y_{k}}$ the expanded version of the latter in the \textit{n}-qubit space affecting qubit $k$, and $CZ$ the 2-qubit corresponding Controlled-Z entanglers indexed by $k$, with target $t$, according to the (graph-stabilizer) subgroup $S_{\text{graph}}<G=SU(2)^{\otimes{n}}$, for the graph with edges $E$ and vertices $V$. From \eqref{eq:16_bis} we notice that parameterization of the fiducial state $|\phi_{\theta}\rangle=V(\theta)|0^{\otimes n}\rangle$ in this case is made dependent on one single scalar $\theta$.   

In our case, we rearrange from \eqref{eq:16} and \eqref{eq:16_bis} resulting into the following data-dependent unitary:

\begin{equation}
\label{eq:17}
U_{COV}(\vec{x})=[\otimes_{k=1}^{n}R_{X}(x_{2k-1})R_{Z}(x_{2k})]\prod_{k=1}^{n}CZ_{k,k+1}
\end{equation}

thus grouping together all $\theta$-independent operations, leaving aside the preparation of the fiducial state (described in more detail in the next section). We also notice from \eqref{eq:17} that the group-specific CZ mapping, originally optimized for the LCE problem, was replaced for a more general \textit{linear} entangling layer. In doing so, we seek to come up with a more general expression that enable us to apply the resulting mapping to a wider range of problems where the possible underlying graph structure is \textit{a priori} unknown. 

\subsubsection{Optimization of the kernel parameters}

We follow the \textit{weighted} kernel alignment procedure described above to fit:

\begin{equation}
\label{eq:18}
\kappa_{\theta}(\vec{x_{i}},\vec{x_{j}})=|\langle0^{\otimes n}|V_{\theta}\textsuperscript{\textdagger}U_{x}\textsuperscript{\textdagger}|U_{x}V_{\theta}|0^{\otimes n}\rangle|^2
\end{equation}

In our experimental setting $U_{x}\in\{U_{ZZ}(\vec{x}), U_{COV}(\vec{x})\}$, respectively from \eqref{eq:15} and \eqref{eq:17}, and $V_{\theta}\in\{V_{shared}(\vec{\theta}), \\V_{dedicated}(\vec{\theta})\}$ where:

\begin{equation}
\label{eq:19}
V_{shared}(\vec{\theta})=\otimes_{k=1}^{n}R_{XYZ}(\theta_{1}, \theta_{2}, \theta_{3})
\end{equation}

and

\begin{equation}
\label{eq:20}
V_{dedicated}(\vec{\theta})= \otimes_{k=1}^{n}R_{XYZ}(\theta_{3(k-1)+1}, \theta_{3(k-1)+2}, \theta_{3(k-1)+3})
\end{equation}

with $R_{XYZ}$ the universal single-qubit rotation gate with 3 Euler angles:

\begin{equation}
\label{eq:21}
R_{XYZ}(\theta_{1}, \theta_{2}, \theta_{3}) =
    \begin{pmatrix}
        \cos\left(\frac{\theta_{1}}{2}\right) & -e^{i\theta_{2}}\sin\left(\frac{\theta_{1}}{2}\right) \\
        e^{i\theta_{3}}\sin\left(\frac{\theta_{1}}{2}\right) & e^{i(\theta_{3}+\theta_{2})}\cos\left(\frac{\theta_{1}}{2}\right)
    \end{pmatrix}  
\end{equation}

One should notice that by choosing $U_{x}=U_{COV}(\vec{x})$ and $V_{shared}(\theta_{1}, \theta_{2}=0, \theta_{3}=0)$ we retrieve the mapping resulting from \eqref{eq:16} and \eqref{eq:16_bis} for a linearly connected graph (for which $R_{XYZ}(\theta, 0, 0) = R_{Y}(\theta)$). Thus, once again, by using \eqref{eq:19} and \eqref{eq:20} we are aiming to generalize the expressivity of the resulting kernel mapping, allowing additional degrees of freedom for adaptability to wider set of classification problems. More specifically, and as the naming suggests, $V_{shared}(\vec{\theta})$ leads to all \textit{n}-qubits in the mapping circuit to share the same parameterized rotation throughout all possible configurations of the Bloch sphere, whereas $V_{dedicated}(\vec{\theta})$ involves a dedicated rotation for each qubit.

\subsection{Benchmark datasets}
Here we describe a benchmark dataset to test the performance of QKE and QKT techniques on diverse classification tasks using the methods and quantum feature mappings described above.  

We proceed to that objective, fully aware of the limitations: gathering a well-diverse and general-enough benchmark dataset that covers all possible situations and enables extraction of unequivocal conclusions in comparing performance between classical and quantum machine learning is a difficult endeavor, if not utopian \cite{bowles_better_2024}. Regardless, despite the challenges, we can still strive for a “fair enough” dataset that captures a broad spectrum of scenarios. The effort is still justified in serving as a valuable resource towards continuing unraveling the potential of quantum computing in machine learning.

With that in mind, we assemble a mix of quantum-inspired artificially-generated data, together with some reference classical machine learning datasets to test versatility and generalization capabilities of both quantum and classical methods in heterogeneous settings.

\subsubsection{Ad-hoc quantum datasets}

An obvious (baseline) choice is to use synthetic ad-hoc data representing classification problems that should be easily separable by the investigated quantum feature mappings, but are theoretically hard for classical methods.

In this regard, we denote by \textit{Ad-hoc-ZZ} dataset, to a 2-dimensional 400 sample of ad-hoc data generated according to the procedure described in \cite{havlicek_supervised_2019} for the \textit{ZZ-FeatureMap}. Data are split in 200 samples corresponding to train and test partitions, each including a balanced representation of the positive and negative classes. Default $\Delta=0.3$ separation gap parameter is used. 

Analogously, we consider the LCE classification problem outlined in the work describing \textit{covariant quantum mappings} \cite{glick_covariant_2024}. More specifically, a 7-qubit instance of the problem is used, which is available online in a related experimental repository on GitHub \cite{quantum-kernel-training}. The resulting \textit{Ad-hoc-COV} dataset comprises 14-dimensional feature vectors with independently sampled training and testing partitions, each consisting of 32 data points per label. 

\subsubsection{Reference classical datasets}
We consider different instances of two well-known datasets in classical machine learning, namely the IRIS and the MNIST classification datasets. We also include a third one, namely MNIST-1D, specially designed low-scale model evaluation. 

First, we denote \textit{Linear-IRIS} the data collection that consists of the first two classes (\textit{setosa} vs \textit{versicolor}) of the famous classification dataset. As the name suggests, these two classes are known to be linearly separable. The resulting dataset involves 100 4-dimensional data samples (50 for each class). We arrange train and test partitions using a 70-30\% stratified split. We also compose a second dataset, referred to as \textit{Non-linear-IRIS} that comprises the \textit{versicolor} vs. \textit{virginica} classification task, which cannot be fully separated without resorting to non-linear feature projections. Similarly, in this case we use a 70-30\% rule for training and test splitting.

From the reference MNIST dataset, we consider two additional instances of a downscaled version of the 3 vs. 5 digits classification task. In particular, we take as reference the 4- and 8-dimensional versions of the $\text{MNIST PCA}^{\_}$ benchmark from \cite{bowles_better_2024}. As the name suggests, generation of these variants involves the application of Principal Component Analysis (PCA) to the original MNIST dataset for the targeted dimensionality reduction, after which a stratified subsample of 250 training and test points is taken to constrain the resulting complexity within the reach range of quantum kernel methods. We denote the two resulting datasets as $\text{MNIST-PCA-4}$ and $\text{MNIST-PCA-8}$, respectively.

Finally, following similar procedures, we arrange two additional 4- and 8-dimensional instance versions using as reference the MNIST-1D dataset, a minimalist, low-memory, and low-compute alternative to the classical MNIST \cite{greydanus2024_mnist1d}. In recent discussion posed at Bowles et al. \cite{bowles_better_2024}, it was suggested that this dataset may be a much more suitable alternative to use for benchmarking of QML models.

Table \ref{tab:datasets} summarizes the characteristics of the datasets included in the resulting benchmark. In order to avoid the potential for bias and to facilitate subsequent statistical analysis, experimental procedures involve multiple repetitions with random resampling of the corresponding training and testing partitions for each dataset.

\begin{table}[]
 \caption{Summary of datasets included in our benchmarking. The number of positive and negative instances in the corresponding training and testing set partitions are shown in the last two columns}
  \centering
  \begin{tabular}{l c c c}
    \toprule
    Dataset & Feature dimension & Training dist. & Test dist. \\
    \midrule
    Ad-hoc-ZZ & 2 & 100 - 100 & 100 - 100 \\
    Ad-hoc-COV & 14 & 32 - 32 & 32 - 32 \\
    Linear-IRIS & 4 & 35 - 35 & 15 - 15 \\
    Non-linear-IRIS & 4 & 35 - 35 & 15 - 15 \\
    MNIST-PCA-4 & 4 & 125 - 125 & 125 - 125 \\
    MNIST-PCA-8 & 8 & 125 - 125 & 125 - 125 \\
    MNIST-1D-PCA-4 & 4 & 125 - 125 & 50 - 50 \\
    MNIST-1D-PCA-8 & 8 & 125 - 125 & 50 - 50 \\
    \bottomrule
  \end{tabular}
  \label{tab:datasets}
\end{table}

\subsubsection{Classical baseline models}

To set a reference with respect to classification performance using quantum kernel methods, we consider a set of classical machine learning counterparts.

For this purpose, we take classical SVMs with the use of \textit{linear}, \textit{polynomial}, and \textit{Radial Basis Function (RBF)} kernels to target the same classification benchmark. We also include a Logistic Regression (LR) classifier for additional reference. Both, the logistic classifier and the linear kernel SVM operate in the original feature space and are limited to the generation of linear decision boundaries. Polynomial and RBF kernels are able to produce non-linear decisions in the induced feature space, as described in Section \ref{subsec:kernel_methods}. For polynomial kernels, it is well-established that the dimensionality of the induced space is of order $\binom{n}{k} = \frac{n!}{k!(n-k)!}$, where \textit{n} is the original data dimension, and \textit{k} is the degree of the polynomial. We shall consider this result when choosing the specific degree of the polynomial for each dataset. RBF-kernels can theoretically map data up to infinite-dimensional feature spaces. However, in practice, the effective dimension is constrained by the number of available training samples \textit{m} \cite{bishop2007}. 

\subsection{Data pre-processing and optimization settings}

For the quantum models, the input features were scaled to the $0-2\pi$ range for all datasets. This is a reasonable choice given that both quantum feature mappings tackled in this work implement \textit{angle encoding} \cite{schuld2021book}. For classical models, we followed the standard approach of removing the mean and scaling features to unit variance (z-score normalization) \cite{bishop2007}. In any case, we should note that the scaling parameters in both cases are learned exclusively from the training set and subsequently applied to the test set, under the assumption that access to the test set during training is not permitted.

We scheduled a hyperparameter search for all models and datasets using a full grid approach. For the classical methods this includes, as a minimum, optimization of the regularization parameter $C$ in the range [0.01, 0.1, 1, 10, 100]. In addition, for the RBF kernel we also explored different configurations for the bandwidth parameter $\gamma \in$ [0.001, 0.001, 0.01, 1, 10]. For classical polynomial kernels, the degree of the polynomial $k$ was set such that the dimensionality of the resulting induced space approximately matches that of the quantum kernels for the specific target dataset. We use the following rule-of-thumb: set $k$ to the closest natural number such that  $\binom{n}{k} \leq 2^{2n}$, i.e., the space of density matrices for a \textit{n}-qubit system (c.f. the mapping \eqref{eq:6}). 

In the case of the quantum kernels, the hyperparameter search includes the same logarithmic progression for $C$ as for classical methods. In addition, we also targeted optimization of the (quantum) kernel bandwidth $\lambda$ exploring the range [0.001, 0.01, 0.1, 0.5, 1.0].
As discussed in related work \cite{Shaydulin_2022, canatar2023bandwidth}, this parameter acts as a trade-off between the generalization and expressiveness capabilities of the quantum kernel, and can play a role in mitigating the effects of quantum kernel concentration \cite{thanasilp2024exponentialconcentrationquantumkernel, miroszewski2024_shots_concentration_QKM}.

Exploration of the hyperparameter space grid was carried out in all cases following a 5-fold cross-validation on the training set partition of the corresponding dataset. We used the accuracy score to select the best model. We note that whereas this metric can be misleading for some classification tasks, especially in the context of multi-class or unbalanced data, we choose it here for its simplicity and clarity of interpretation, and since our datasets are all binary and balanced. Once the best hyperparameter set has been decided for each model, generalization performance is evaluated in the corresponding testing set. More generally, to quantify the performance of each investigated classifier, in addition to the accuracy score, Cohen's kappa ($\kappa$) and macro F1-score were also calculated. 

For QKT optimization, we allowed a maximum of 400 iterations of the associated SPSA hybrid algorithm discussed in the methods section. We used a second-order SPSA version that includes Hessian estimation of the loss for improved convergence. We also resorted to automatic calibration to avoid uninformed fixing of the learning rate and the perturbation parameter. As starting point, we initialized the parameter vector $\vec{\theta}$ to all zeros. We chose this point as it corresponds with the absence of effective rotation applied by means of $V_{\theta}$ with respect to the $|0^{\otimes n}\rangle$ state vector initialization. See \eqref{eq:18} with regard to the associated quantum kernel estimation procedure. Once the maximum number of allowed iterations is reached, $\vec{\theta_{opt}}$ is finally set to the point with corresponding minimal loss (maximal weighted kernel alignment) obtained during the training phase.    

All experiments were carried out in Python using Qiskit \cite{qiskit2024} (v1.1.0) in combination with \textit{scikit-learn} \cite{scikit-learn} (v1.4.2.) packages. The latter was used for the implementation of the classical models, general hyperparameter tuning, metric evaluations, and to solve the SVM optimization problem. For QKT optimization, we relied on the SPSA implementation provided by the \textit{qiskit-algorithms} (v0.3.0) extension package. For estimation of the quantum kernel matrices for QKE, we used statevector simulations carried out on a classical computer. For this purpose, we used functionality of the \verb|TrainableFidelityStatevectorKernel| and \verb|QuantumKernelTrainer| classes contained in the \textit{qiskit-machine-learning} (v0.7.2) package. The source code for the reproducibility of all the experimental procedures in this study is available on GitHub (\url{https://github.com/diegoalvareze/qkt_benchmarking}).

\section{Results}
\label{sec:results}

Experimental simulations for each model and dataset were repeated a total of $n=30$ times. For each repetition, the corresponding training and testing partitions were regenerated using random resampling following the distributions outlined in \cref{tab:datasets}. This was in order to minimize potential biases and to enable subsequent statistical analyses of the results.

\cref{tab:table2_quantum_methods} shows the corresponding performance values obtained for the quantum models. In the table, the first column describes the dataset, the second column describes the applied quantum feature map, and the third column refers to the QKT parameterization strategy. Rows without an entry description under this category signify the absence of kernel training, indicating the implementation of "plain" QKE. Otherwise either the text \textit{shared} or \textit{dedicated} is shown, c.f. \eqref{eq:19} and \eqref{eq:20} respectively, together with the number of involved parameters. Notice that this number depends, not only on the followed parameterization strategy, but also on the circuit layout that results from the application of the associated quantum feature map. The subsequent columns show the corresponding averaged and standard deviation results for each of the evaluated metrics, obtained at training and testing times, respectively. Performance results obtained by the analyzed classical methods on each case are shown in \cref{tab:table_classical_methods}.

\begin{table*}[]
 \caption{Experimental results for quantum methods}
  \centering
  \resizebox{\textwidth}{!}{%
  \begin{tabular}{l l l c c c c c c}
    \toprule
    & & QKT approach & & & & & & \\
    Dataset & Mapping & (\# parameters) & $ACC_{TR}$ & $\kappa_{TR}$ & $MF_{TR}$ & $ACC_{TS}$ & $\kappa_{TS}$ & $MF_{TS}$ \\
    \midrule
    Ad-hoc-ZZ & \textbf{ZZFeatureMap} & \textbf{---} & \textbf{0.99 (0.01)} & \textbf{0.99 (0.03)} & \textbf{0.99 (0.01)} & \textbf{0.99 (0.02)} & \textbf{0.98 (0.03)} & \textbf{0.99 (0.02)} \\
    & & \textbf{Shared (3)} & \textbf{0.99 (0.01)} & \textbf{0.99 (0.03)} & \textbf{0.99 (0.01)} & \textbf{0.99 (0.01)} & \textbf{0.98 (0.03)} & \textbf{0.99 (0.01)} \\
    & & \textbf{Dedicated (6)} & \textbf{0.99 (0.01)} & \textbf{0.99 (0.03)} & \textbf{0.99 (0.01)} & \textbf{0.99 (0.02)} & \textbf{0.98 (0.05)} & \textbf{0.99 (0.02)} \\
    & CovariantMap & --- & 0.60 (0.03) & 0.19 (0.07) & 0.59 (0.03) & 0.57 (0.04) & 0.15 (0.08) & 0.57 (0.04) \\
    & & Shared (3) & 0.64 (0.05) & 0.29 (0.10) & 0.64 (0.05) & 0.61 (0.07) & 0.22 (0.14) & 0.61 (0.07) \\
    & & Dedicated (3) & 0.64 (0.05) & 0.29 (0.10) & 0.64 (0.05) & 0.61 (0.07) & 0.22 (0.14) & 0.61 (0.07) \\
    \midrule
    Ad-hoc-COV & ZZFeatureMap & --- & \textbf{0.99 (0.01)} & \textbf{0.99 (0.02)} & \textbf{0.99 (0.01)} & 0.92 (0.07) & 0.84 (0.14) & 0.92 (0.07) \\
    & & Shared (3) & 0.77 (0.07) & 0.55 (0.14) & 0.77 (0.07) & 0.72 (0.07) & 0.44 (0.13) & 0.71 (0.07) \\
    & & Dedicated (42) & 0.79 (0.07) & 0.58 (0.14) & 0.79 (0.07) & 0.74 (0.04) & 0.48 (0.09) & 0.73 (0.04) \\
    & \textbf{CovariantMap} & --- & 0.97 (0.02) & 0.94 (0.05) & 0.97 (0.02) & 0.88 (0.03) & 0.76 (0.06) & 0.88 (0.03) \\
    & & Shared (3) & \textbf{0.99 (0.02)} & 0.97 (0.04) & \textbf{0.99 (0.02)} & 0.91 (0.03) & 0.81 (0.06) & 0.91 (0.03) \\
    & & \textbf{Dedicated (21)} & \textbf{0.99 (0.01)} & 0.98 (0.02) & \textbf{0.99 (0.01)} & \textbf{0.94 (0.04)} & \textbf{0.88 (0.08)} & \textbf{0.94 (0.04)} \\
    \midrule
    Linear-IRIS & \textbf{ZZFeatureMap} & \textbf{---} & \textbf{1.00 (0.00)} & \textbf{1.00 (0.00)} & \textbf{1.00 (0.00)} & \textbf{1.00 (0.00)} & \textbf{1.00 (0.00)} & \textbf{1.00 (0.00)} \\
    & & Shared (3) & \textbf{1.00 (0.00)} & \textbf{1.00 (0.01)} & \textbf{1.00 (0.00)} & 0.99 (0.02) & 0.99 (0.05) & 0.99 (0.03) \\
    & & \textbf{Dedicated (12)} & \textbf{1.00 (0.00)} & \textbf{1.00 (0.01)} & \textbf{1.00 (0.00)} & \textbf{1.00 (0.02)} & \textbf{1.00 (0.05)} & \textbf{1.00 (0.03)} \\
    & \textbf{CovariantMap} & --- & \textbf{1.00 (0.00)} & \textbf{1.00 (0.00)} & \textbf{1.00 (0.00)} & 0.99 (0.02) & 0.99 (0.05) & 0.99 (0.03) \\
    & & Shared (3) & \textbf{1.00 (0.00)} & \textbf{1.00 (0.00)} & \textbf{1.00 (0.00)} & 0.99 (0.02) & 0.99 (0.05) & 0.99 (0.03) \\
    & & \textbf{Dedicated (6)} & \textbf{1.00 (0.00)} & \textbf{1.00 (0.00)} & \textbf{1.00 (0.00)} & \textbf{1.00 (0.00)} & \textbf{1.00 (0.00)} & \textbf{1.00 (0.00)} \\
    \midrule
    Non-linear-IRIS & ZZFeatureMap & --- & 0.95 (0.02) & 0.90 (0.03) & 0.95 (0.02) & 0.93 (0.03) & 0.85 (0.06) & 0.93 (0.03) \\
    & & Shared (3) & 0.95 (0.02) & 0.90 (0.04) & 0.95 (0.02) & 0.93 (0.03) & 0.85 (0.06) & 0.93 (0.03) \\
    & & Dedicated (12) & 0.96 (0.02) & 0.93 (0.03) & 0.96 (0.02) & 0.93 (0.04) & 0.86 (0.08) & 0.93 (0.04) \\
    & \textbf{CovariantMap} & --- & 0.96 (0.02) & 0.92 (0.03) & 0.96 (0.02) & 0.92 (0.04) & 0.84 (0.07) & 0.92 (0.04) \\
    & & \textbf{Shared (3)} & \textbf{0.97 (0.02)} & \textbf{0.94 (0.03)} & \textbf{0.97 (0.02)} & \textbf{0.94 (0.05)} & \textbf{0.88 (0.09)} & \textbf{0.94 (0.05)} \\
    & & \textbf{Dedicated (6)} & 0.96 (0.02) & 0.93 (0.04) & 0.96 (0.02) & \textbf{0.94 (0.04)} & \textbf{0.88 (0.09)} & \textbf{0.94 (0.04)} \\
    \midrule
    MNIST-PCA-4 & \textbf{ZZFeatureMap} & \textbf{---} & 0.83 (0.04) & \textbf{0.67 (0.09)} & \textbf{0.83 (0.04)} & \textbf{0.80 (0.04)} & \textbf{0.61 (0.07)} & \textbf{0.80 (0.04)} \\
    & & \textbf{Shared (3)} & 0.83 (0.07) & 0.66 (0.13) & \textbf{0.83 (0.08)} & \textbf{0.80 (0.06)} & 0.60 (0.12) & 0.79 (0.07) \\
    & & \textbf{Dedicated (12)} & \textbf{0.84 (0.05)} & \textbf{0.67 (0.11)} & \textbf{0.83 (0.06)} & \textbf{0.80 (0.04)} & 0.60 (0.08) & \textbf{0.80 (0.04)} \\
    & CovariantMap & --- & 0.69 (0.02) & 0.38 (0.03) & 0.69 (0.02) & 0.69 (0.03) & 0.38 (0.06) & 0.68 (0.03) \\
    & & Shared (3) & 0.71 (0.04) & 0.41 (0.07) & 0.70 (0.04) & 0.71 (0.04) & 0.42 (0.08) & 0.70 (0.05) \\
    & & Dedicated (6) & 0.70 (0.03) & 0.40 (0.06) & 0.70 (0.03) & 0.70 (0.03) & 0.39 (0.07) & 0.69 (0.04) \\
    \midrule
    MNIST-PCA-8 & \textbf{ZZFeatureMap} & \textbf{---} & \textbf{0.97 (0.01)} & 0.94 (0.03) & \textbf{0.97 (0.01)} & \textbf{0.95 (0.01)} & \textbf{0.89 (0.02)} & \textbf{0.95 (0.01)} \\
    & & \textbf{Shared (3)} & \textbf{0.97 (0.01)} & \textbf{0.95 (0.03)} & \textbf{0.97 (0.01)} & \textbf{0.95 (0.01)} & \textbf{0.89 (0.03)} & \textbf{0.95 (0.01)} \\
    & & Dedicated (24) & \textbf{0.97 (0.01)} & \textbf{0.95 (0.02)} & \textbf{0.97 (0.01)} & 0.94 (0.02) & \textbf{0.89 (0.03)} & 0.94 (0.02) \\
    & CovariantMap & --- & 0.91 (0.01) & 0.81 (0.03) & 0.91 (0.01) & 0.88 (0.02) & 0.76 (0.05) & 0.88 (0.02) \\
    & & Shared (3) & 0.92 (0.03) & 0.84 (0.05) & 0.92 (0.03) & 0.89 (0.03) & 0.78 (0.05) & 0.89 (0.03) \\
    & & Dedicated (12) & 0.92 (0.03) & 0.83 (0.05) & 0.92 (0.03) & 0.89 (0.02) & 0.78 (0.05) & 0.89 (0.02) \\
    \midrule
    MNIST-1D-PCA-4 & \textbf{ZZFeatureMap} & \textbf{---} & \textbf{0.89 (0.03)} & 0.78 (0.06) & \textbf{0.89 (0.03)} & 0.85 (0.04) & 0.70 (0.08) & 0.85 (0.04) \\
    & & \textbf{Shared (3)} & \textbf{0.89 (0.03)} & 0.78 (0.06) & \textbf{0.89 (0.03)} & 0.85 (0.04) & \textbf{0.71 (0.08)} & 0.85 (0.04) \\
    & & \textbf{Dedicated (12)} & \textbf{0.89 (0.03)} & \textbf{0.79 (0.06)} & \textbf{0.89 (0.03)} & \textbf{0.86 (0.04)} & \textbf{0.71 (0.07)} & \textbf{0.86 (0.04)} \\
    & CovariantMap & --- & 0.74 (0.05) & 0.47 (0.10) & 0.73 (0.05) & 0.71 (0.06) & 0.41 (0.11) & 0.70 (0.06) \\
    & & Shared (3) & 0.76 (0.05) & 0.51 (0.10) & 0.75 (0.05) & 0.73 (0.06) & 0.46 (0.11) & 0.73 (0.06) \\
    & & Dedicated (6) & 0.76 (0.05) & 0.52 (0.10) & 0.76 (0.05) & 0.74 (0.06) & 0.47 (0.12) & 0.73 (0.06) \\
    \midrule
    MNIST-1D-PCA-8 & \textbf{ZZFeatureMap} & \textbf{---} & 0.95 (0.02) & 0.90 (0.05) & 0.95 (0.02) & \textbf{0.92 (0.03)} & \textbf{0.84 (0.06)} & \textbf{0.92 (0.03)} \\
    & & \textbf{Shared (3)} & 0.95 (0.02) & 0.90 (0.05) & 0.95 (0.02) & \textbf{0.92 (0.03)} & \textbf{0.84 (0.06)} & \textbf{0.92 (0.03)} \\
    & & \textbf{Dedicated (24)} & \textbf{0.96 (0.02)} & \textbf{0.92 (0.04)} & \textbf{0.96 (0.02)} & \textbf{0.92 (0.03)} & \textbf{0.84 (0.06)} & \textbf{0.92 (0.03)} \\
    & CovariantMap & --- & 0.79 (0.02) & 0.57 (0.05) & 0.78 (0.02) & 0.80 (0.03) & 0.60 (0.07) & 0.80 (0.03) \\
    & & Shared (3) & 0.81 (0.05) & 0.61 (0.10) & 0.80 (0.05) & 0.81 (0.05) & 0.63 (0.09) & 0.81 (0.05) \\
    & & Dedicated (12) & 0.80 (0.05) & 0.61 (0.10) & 0.80 (0.05) & 0.81 (0.05) & 0.63 (0.09) & 0.81 (0.05) \\
    \bottomrule
    \multicolumn{9}{l}{Average (standard deviation) values for all $n=30$ repetitions are reported; ACC = accuracy, $\kappa$ = Cohen's kappa, MF = Macro F1-score;}\\
    \multicolumn{9}{l}{Best values for each metric are highlighted in bold. Best model is marked as the one maximizing testing accuracy.}\\
  \end{tabular}}
  \label{tab:table2_quantum_methods}
\end{table*}

\begin{table*}[]
 \caption{Results of classical methods}
  \centering
  \resizebox{\textwidth}{!}{%
  \begin{tabular}{l l c c c c c c}
    \toprule
    Dataset & Model & $ACC_{TR}$ & $\kappa_{TR}$ & $MF_{TR}$ & $ACC_{TS}$ & $\kappa_{TS}$ & $MF_{TS}$ \\
    \midrule
    Ad-hoc-ZZ & LR & 0.55 (0.03) & 0.09 (0.06) & 0.55 (0.03) & 0.51 (0.04) & 0.01 (0.09) & 0.50 (0.04) \\
    & SVM linear & 0.55 (0.03) & 0.10 (0.06) & 0.55 (0.03) & 0.51 (0.04) & 0.02 (0.07) & 0.51 (0.04) \\
    & SVM poly (d=4) & 0.57 (0.03) & 0.13 (0.07) & 0.52 (0.06) & 0.54 (0.03) & 0.08 (0.05) & 0.49 (0.05) \\
    & \textbf{SVM rbf} & \textbf{0.99 (0.02)} & \textbf{0.98 (0.05)} & \textbf{0.99 (0.02)} & \textbf{0.87 (0.04)} & \textbf{0.74 (0.09)} & \textbf{0.87 (0.04)} \\
    \midrule
    Ad-hoc-COV & \textbf{LR} & \textbf{0.98 (0.02)} & \textbf{0.97 (0.04)} & \textbf{0.98 (0.02)} & \textbf{0.90 (0.04)} & \textbf{0.80 (0.09)} & \textbf{0.90 (0.04)} \\
    & \textbf{SVM linear} & \textbf{0.98 (0.02)} & \textbf{0.97 (0.04)} & \textbf{0.98 (0.02)} & \textbf{0.90 (0.05)} & 0.79 (0.11) & \textbf{0.90 (0.05)} \\
    & SVM poly (d=17) & 0.69 (0.08) & 0.38 (0.16) & 0.65 (0.11) & 0.50 (0.00) & 0.00 (0.00) & 0.34 (0.01) \\
    & SVM rbf & 0.97 (0.07) & 0.93 (0.13) & 0.97 (0.07) & 0.83 (0.07) & 0.67 (0.15) & 0.83 (0.08) \\
    \midrule
    Linear-IRIS & \textbf{LR} & \textbf{1.00 (0.00)} & \textbf{1.00 (0.00)} & \textbf{1.00 (0.00)} & \textbf{1.00 (0.00)} & \textbf{1.00 (0.00)} & \textbf{1.00 (0.00)} \\
    & \textbf{SVM linear} & \textbf{1.00 (0.00)} & \textbf{1.00 (0.00)} & \textbf{1.00 (0.00)} & \textbf{1.00 (0.00)} & \textbf{1.00 (0.00)} & \textbf{1.00 (0.00)} \\
    & SVM poly (d=7) & 0.99 (0.01) & 0.97 (0.03) & 0.99 (0.01) & 0.97 (0.03) & 0.94 (0.06) & 0.97 (0.03) \\
    & \textbf{SVM rbf} & \textbf{1.00 (0.00)} & \textbf{1.00 (0.00)} & \textbf{1.00 (0.00)} & \textbf{1.00 (0.00)} & \textbf{1.00 (0.00)} & \textbf{1.00 (0.00)} \\
    \midrule
    Non-linear-IRIS & \textbf{LR} & 0.97 (0.02) & 0.93 (0.04) & 0.97 (0.02) & \textbf{0.94 (0.04)} & \textbf{0.88 (0.07)} & \textbf{0.94 (0.04)} \\
    & \textbf{SVM linear} & 0.97 (0.02) & 0.94 (0.03) & 0.97 (0.02) & \textbf{0.94 (0.03)} & \textbf{0.88 (0.06)} & \textbf{0.94 (0.03)} \\
    & SVM poly (d=7) & 0.84 (0.03) & 0.69 (0.05) & 0.84 (0.03) & 0.81 (0.05) & 0.62 (0.11) & 0.80 (0.06) \\
    & \textbf{SVM rbf} & \textbf{0.98 (0.02)} & \textbf{0.95 (0.04)} & \textbf{0.98 (0.02)} & \textbf{0.94 (0.04)} & \textbf{0.88 (0.08)} & \textbf{0.94 (0.04)} \\
    \midrule
    MNIST-PCA-4 & LR & 0.78 (0.02) & 0.56 (0.04) & 0.78 (0.02) & 0.78 (0.03) & 0.56 (0.05) & 0.78 (0.03) \\
    & SVM linear & 0.78 (0.02) & 0.56 (0.03) & 0.78 (0.02) & 0.78 (0.03) & 0.56 (0.06) & 0.78 (0.03) \\
    & SVM poly (d=7) & \textbf{0.87 (0.04)} & \textbf{0.73 (0.08)} & \textbf{0.87 (0.04)} & 0.75 (0.04) & 0.50 (0.08) & 0.74 (0.04) \\
    & \textbf{SVM rbf} & 0.85 (0.03) & 0.69 (0.06) & 0.85 (0.03) & \textbf{0.82 (0.03)} & \textbf{0.65 (0.06)} & \textbf{0.82 (0.03)} \\
    \midrule
    MNIST-PCA-8 & LR & 0.93 (0.02) & 0.86 (0.03) & 0.93 (0.02) & 0.93 (0.02) & 0.86 (0.03) & 0.93 (0.02) \\
    & SVM linear & 0.93 (0.02) & 0.87 (0.04) & 0.93 (0.02) & 0.92 (0.02) & 0.85 (0.03) & 0.92 (0.02) \\
    & SVM poly (d=11) & 0.86 (0.04) & 0.72 (0.07) & 0.86 (0.04) & 0.77 (0.05) & 0.53 (0.09) & 0.75 (0.05) \\
    & \textbf{SVM rbf} & \textbf{0.98 (0.01)} & \textbf{0.97 (0.03)} & \textbf{0.98 (0.01)} & \textbf{0.95 (0.01)} & \textbf{0.89 (0.03)} & \textbf{0.95 (0.01)} \\
    \midrule
    MNIST-1D-PCA-4 & LR & 0.82 (0.02) & 0.65 (0.04) & 0.82 (0.02) & 0.82 (0.03) & 0.64 (0.07) & 0.82 (0.03) \\
    & SVM linear & 0.83 (0.02) & 0.65 (0.04) & 0.83 (0.02) & 0.82 (0.03) & 0.65 (0.07) & 0.82 (0.03) \\
    & SVM poly (d=7) & \textbf{0.92 (0.03)} & \textbf{0.85 (0.05)} & \textbf{0.92 (0.03)} & 0.74 (0.04) & 0.49 (0.08) & 0.73 (0.05) \\
    & \textbf{SVM rbf} & 0.89 (0.03) & 0.79 (0.06) & 0.89 (0.03) & \textbf{0.87 (0.04)} & \textbf{0.74 (0.08)} & \textbf{0.87 (0.04)} \\
    \midrule
    MNIST-1D-PCA-8 & LR & 0.87 (0.02) & 0.73 (0.03) & 0.87 (0.02) & 0.87 (0.03) & 0.73 (0.07) & 0.87 (0.03) \\
    & SVM linear & 0.87 (0.02) & 0.73 (0.03) & 0.87 (0.02) & 0.87 (0.03) & 0.74 (0.06) & 0.87 (0.03) \\
    & SVM poly (d=11) & \textbf{0.96 (0.01)} & \textbf{0.92 (0.02)} & \textbf{0.96 (0.01)} & 0.70 (0.03) & 0.41 (0.06) & 0.68 (0.04) \\
    & \textbf{SVM rbf} & \textbf{0.96 (0.03)} & \textbf{0.92 (0.06)} & \textbf{0.96 (0.03)} & \textbf{0.93 (0.03)} & \textbf{0.86 (0.05)} & \textbf{0.93 (0.03)} \\
    \bottomrule
    \multicolumn{8}{l}{Average (standard deviation) values for all $n=30$ repetitions are reported;}\\
    \multicolumn{8}{l}{ACC = accuracy, $\kappa$ = Cohen's kappa, MF = Macro F1-score;}\\
    \multicolumn{8}{l}{Best values for each metric are highlighted in bold. Best model is marked as the one maximizing testing accuracy.}\\
  \end{tabular}}
  \label{tab:table_classical_methods}
\end{table*}

\begin{figure*}[]
\centerline{\includegraphics[width=0.9\textwidth]{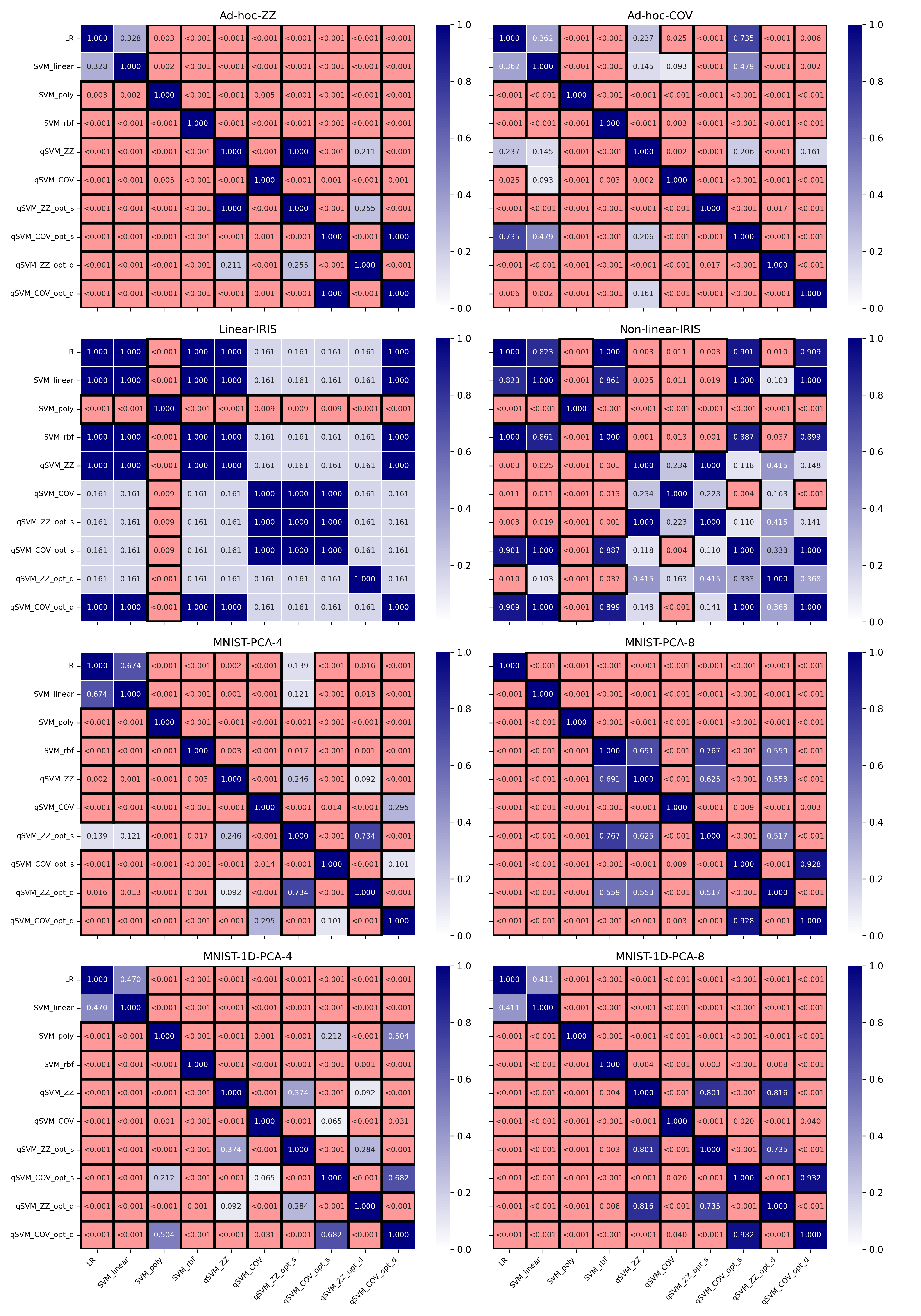}}
\caption{P-values for t-test comparing paired classification accuracy scores in the testing set for each combination of tested models and for each dataset}
\label{fig:significance_TS}
\end{figure*}

To analyze the statistical significance of the differences found across the models, we executed paired t-tests among the corresponding metrics distributions. The results are shown in \cref{fig:significance_TS}, where we take as reference the classification accuracy achieved in the testing set. 

\begin{figure*}[]
\centerline{\includegraphics[width=0.9\textwidth]{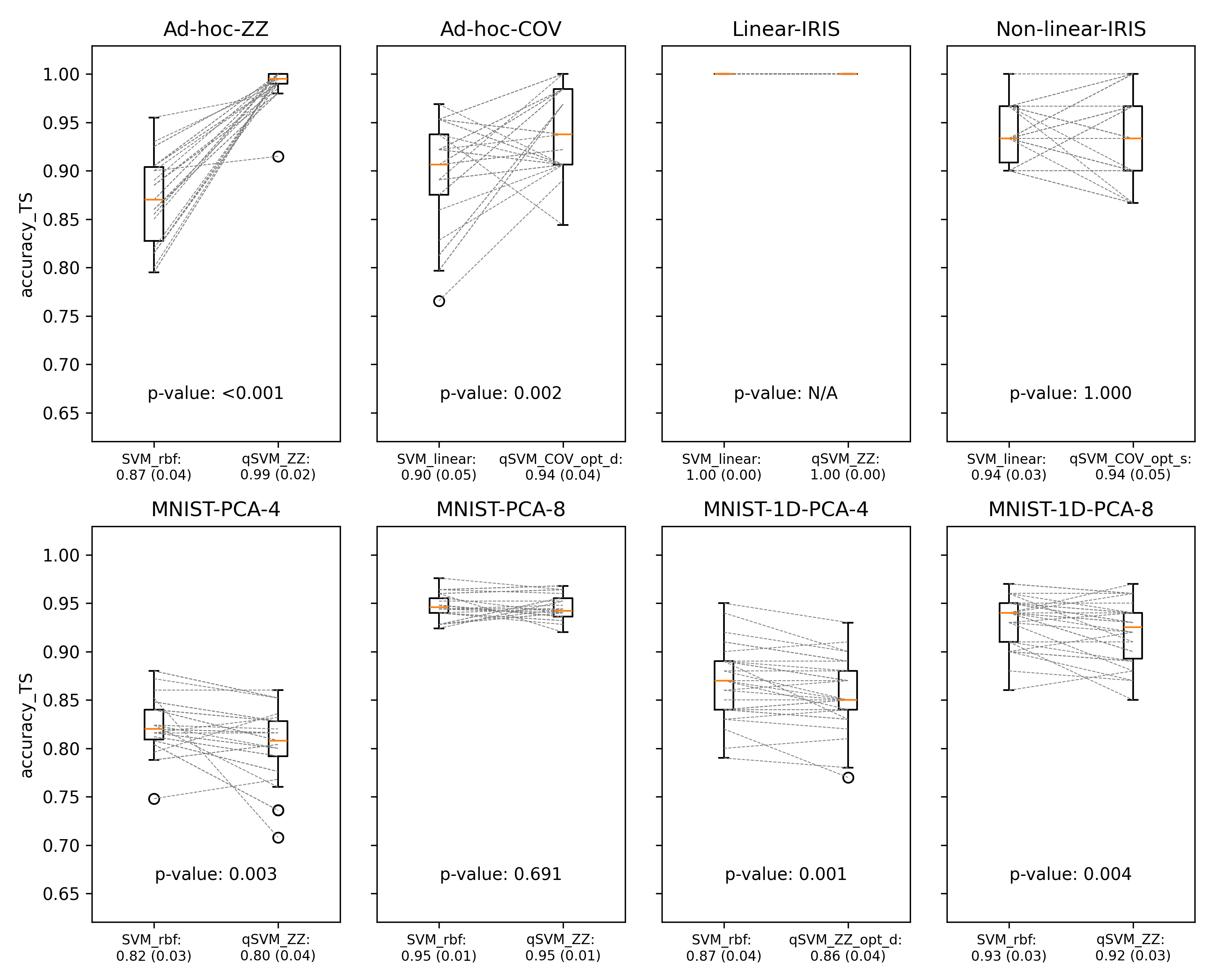}}
\caption{Comparison of the best classical and quantum models per dataset among $n=30$ repetitions.}
\label{fig:best_classical_vs_quantum}
\end{figure*}

A more detailed comparison between the respective classical and quantum methods that perform best on each dataset is shown in \cref{fig:best_classical_vs_quantum}. We see that quantum methods outperform classical models in the subset of ad-hoc datasets, but not in the general case of classical problems. In some cases (Linear-IRIS, Non-linear-IRIS, MNIST-PCA-8), quantum methods are able to produce results that are statistically indistinguishable from classical ones. It is also noted that both approaches are capable of achieving perfect class separation in the Linear-IRIS. However, in others (MNIST-PCA-4, MNIST-1D-PCA-4, MNIST-PCA-1D-8), a significant slight degradation in performance is observed.

The experimental data suggest an overall better performance of quantum methods based on variants of the ZZ-mapping over the Covariant alternative. Apart from the otherwise logical trend observed in the respective ad-hoc datasets, variants of the ZZ-mapping achieve the best performance in four out of the six classical datasets. However, in the two exceptions (Linear-IRIS and Non-linear-IRIS), it follows from \cref{fig:significance_TS} that the differences between the two quantum mappings do not reach the level of statistical significance.

\begin{figure*}[]
\centerline{\includegraphics[width=0.9\textwidth]{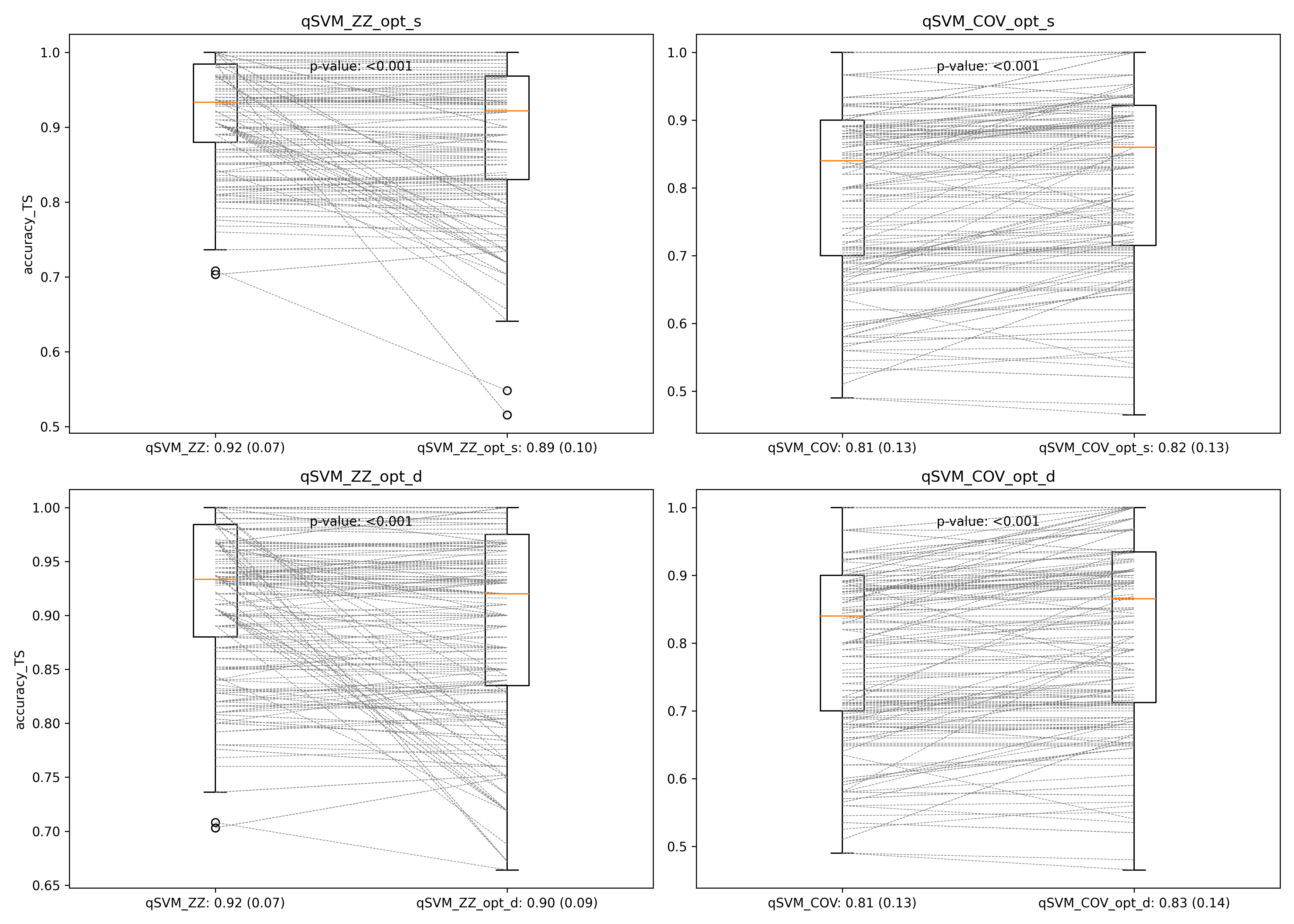}}
\caption{Differences in test accuracy score between baseline QKE and after the quantum kernel was optimized using QKT with weighted alignment. Performance values are aggregated across all datasets. P-values for paired t-tests between the corresponding metric samples are superimposed on each plot.}
\label{fig:qkt_perf_contrib}
\end{figure*}

To provide a general evaluation of the contribution of the weighted alignment QKT with respect to baseline QKE, as well as to analyze the influence of the associated parameterization strategy, aggregation of the corresponding performance metrics is considered across all datasets. The results are shown in \cref{fig:qkt_perf_contrib}. We observe an opposite trend depending on the specific quantum mapping. Whereas for the Covariant mapping QKT seems to contribute to an overall improvement, regardless of the parameterization strategy, a downgrading effect is observed in the case of the ZZ feature map. The statistical significance of the results at the dataset levels shows mixed behavior.

\begin{figure*}[]
\centerline{\includegraphics[width=0.9\textwidth]{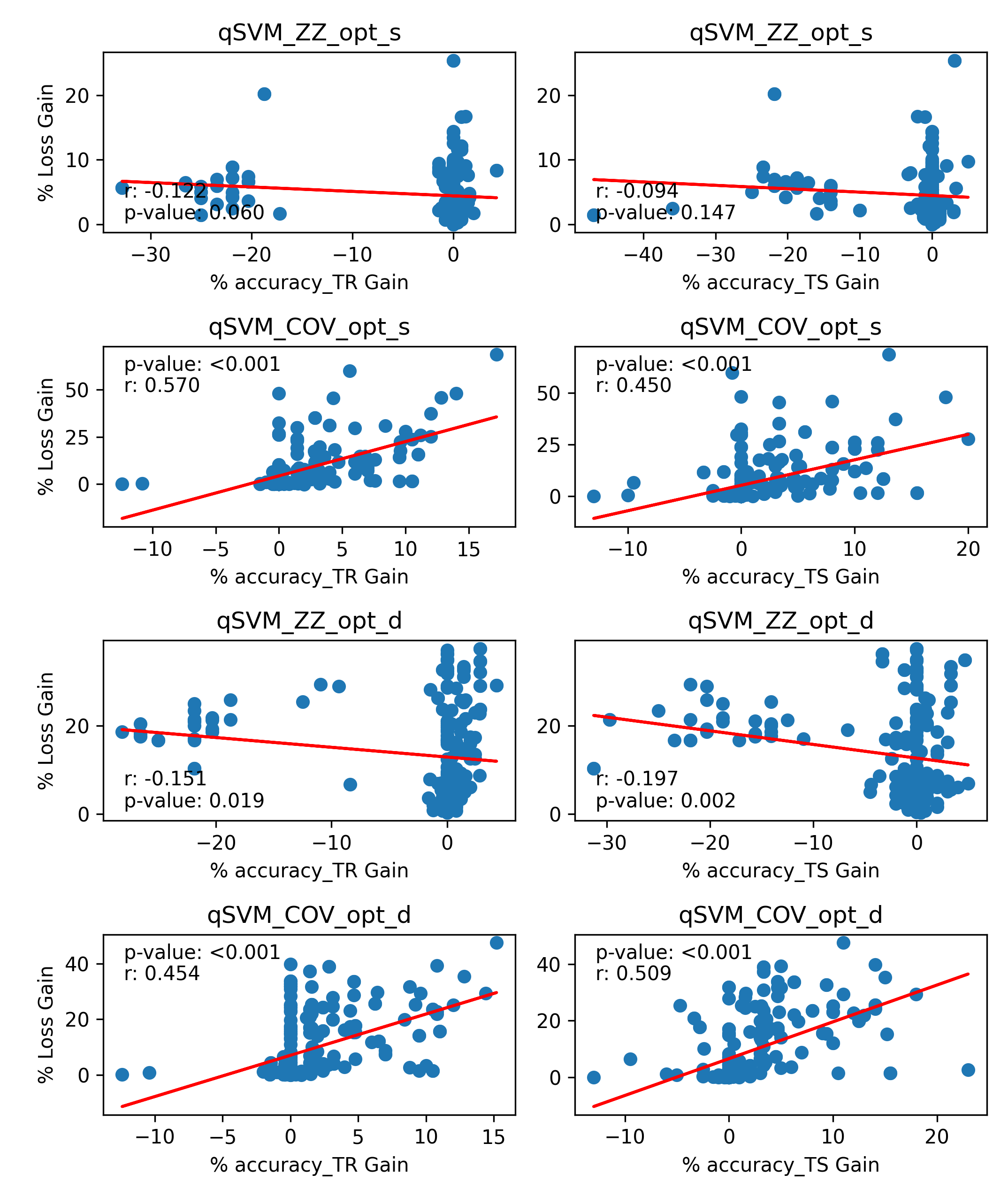}}
\caption{Relationship between evolution of the target loss and resulting accuracy score.}
\label{fig:qkt_loss_perf_relationship}
\end{figure*}

We can further analyze the relationship between the minimization of the target loss during the training phase and the resulting performance gain. The results in terms of the respective training and testing classification accuracies are shown in \cref{fig:qkt_loss_perf_relationship}. The data confirm the trends of \cref{fig:qkt_perf_contrib}, observing that while the achieved improvement in the target loss goes up to 50\% in some cases, it does not always translate into an effective increase in the resulting classification performance. Furthermore, a general downgrading effect can be observed for the ZZ feature map, and even in some specific cases when using Covariant kernels (negative values on the x-axis). Nevertheless, general positive correlation between the weighted target alignment and the classification performance is observed for the latter.

\section{Discussion and conclusions}
\label{sec:discussion}

Following the results of our experiments, one initial observation is that there is no one-fits-to-all method; the best approach in terms of classification performance, whether quantum or classical (with their corresponding variants) depends very much on the specific targeted dataset. Perhaps unsurprisingly, quantum methods outperformed classical alternatives in the Ad-hoc-ZZ and Ad-hoc-COV datasets. More specifically, the optimal points for these datasets were achieved when using the corresponding quantum generative mappings, ZZFeatureMap and CovariantMap, respectively. This is consistent with the expected behavior predicted in the original works \cite{havlicek_supervised_2019, glick_covariant_2024}. However, these works did not use classical methods to compare performance on the corresponding ad-hoc data. Thus, our results confirm these hypotheses with experiments. The \textit{quantum suitability} nevertheless vanishes for the subset of classical datasets. This does not imply that quantum methods were ineffective or unable to learn. With some exceptions (MNIST-PCA-4, MNIST-1D-PCA-4, and MNIST-1D-PCA-8), they achieved results that were statistically indistinguishable from the best classical alternatives. And even then, the average performance scores were not far off from those of the classical methods. 

In terms of the quantum mapping, our data suggest better overall performance in the case of the ZZ-mapping variant compared to the Covariant counterparts. However, it is important to recall that in order to extend the use of the Covariant mapping beyond the specific LCE problem, the actual optimal coupling for the Ad-hoc-COV data was replaced by a more general linear entanglement layer. This modification comes with the cost of preventing the mapping from achieving perfect classification in the corresponding Ad-hoc-COV dataset. The extent to which the differences in entanglement levels between the two mappings may have conditioned their overall performance on the classical datasets remains a topic for further research.

Perhaps relatedly, when analyzing the effect of QKT over baseline QKE, we observed opposite trends between the two quantum mappings. Whereas QKT seemed to contribute to an overall slight improvement for the Covariant mapping, a downgrading effect was observed for the ZZ- feature maps. One could speculate on whether part of this trend could be explained by the fact that the room for improvement with respect to baseline QKE was more restricted for the ZZFeatureMap. However, the data in \cref{fig:qkt_loss_perf_relationship} seem to discard possible overfitting effects.

As for the influence of the parameterization strategy, our data show that the \textit{dedicated} approach always achieves equal or better performance than the \textit{shared} strategy. Although these differences do not always reach the level of statistical significance, it is a logical result for which the space of \textit{shared} solutions is an obvious subset of the \textit{dedicated} parameter vectors. That is, where all dedicated rotations share exactly the same parameter values. 

Altogether, the results of our benchmarking call into question a generalized added value of applying QKT optimization. As discussed later, the additional computational cost associated with QKT optimization is significant, and without a clear benefit in terms of performance gain, it is difficult to justify its use.

In fact, efforts toward optimization of the kernel hyperparameters may pay off more effectively. As a reference, one can compare the results of the present study with preliminary work that relied on fixed default parameterization \cite{alvarezestevez2024benchmark_v1}, noting significant changes in the simulation results that boost the resulting performance of both classical and quantum approaches. Notably, in addition to the regularization parameter $C$, this study included the quantum kernel bandwidth in the hyperparameter search. As discussed in related work \cite{Shaydulin_2022, canatar2023bandwidth}, this parameter acts as a trade-off between the generalization and expressiveness capabilities of the quantum kernel, and can play a role in mitigating the effects of quantum kernel concentration. To the best of our knowledge, this is the first study to consider this parameterization in conjunction with quantum kernel training. Although an explicit analysis of the relationship between quantum kernel alignment and optimal bandwidth is beyond the scope of this work, it motivates further studies to deeply understand the underlying implications of this interplay.

From a general point of view, it is also important to consider that quantum feature maps can be constructed with considerable freedom regarding their architecture. In this work, we have chosen ZZFeatureMap and CovariantMaps as a basis, for which their conjectured association with quantum advantage. Whereas our primary focus targeted quantum kernel optimization, rather than kernel selection \cite{hubregtsen_training_2022}, one still has to address the problem of selecting the kernel family (e.g., the structure of the parameterized circuit layout). One should expect this choice to largely influence the foreseeable performance on the targeted dataset, potentially more so than by (QKT) parameter optimization.

From a non-functional point of view, when comparing classical vs. quantum kernel methods, one should also consider important aspects that go beyond the mere performance metrics. These include issues related to their scalability, efficiency, robustness, or practical usability. In this regard, kernel methods (including classical ones) need to store and compute the kernel matrix, which grows quadratically with the number of training samples. Hence, these methods become computationally expensive as the dataset size increases. In fact, in the current era of big data, kernel methods have been superseded by (deep) neural networks, which usually scale linearly on the number of training samples. This is without prejudice to other advantages that kernel-based methods bring to (still challenging) problems of a more moderated size. These include improved efficiency, interpretability or convergence guarantees, partly derived from the well-defined mathematical framework supporting these methods.

However, in the case of quantum kernels, the scalability aspect becomes even more crucial. Specifically, current practical implementation of the quantum fidelity estimation procedure requires multiple repetitions of the compute-uncompute circuit with increased cost penalty over classical counterparts \cite{havlicek_supervised_2019}. Moreover, variational QKT optimization involves additional computational overhead over QKE, for which each iteration of the hybrid algorithm requires re-computation of the kernel matrix with the new set of candidate parameters. In addition, one should take into account current NISQ-era hardware limitations, that make it necessary to increase the number of sampling shots to compensate for the noise effects. Recent work has also discussed the related problem of exponential concentration, whereby values of the kernel matrix tend to concentrate around constant values, and for which precision of the related estimated entries needs to be increased to allow learning to occur\cite{thanasilp2024exponentialconcentrationquantumkernel, miroszewski2024_shots_concentration_QKM}. 

Whether these issues may potentially invalidate the theoretical exponential speedups attributed to the use of quantum kernels is still under debate. Recent work has explored alternative approaches, such as the use of projected kernels \cite{Huang_2021}, or the optimization of the number of shots \cite{miroszewski2024_shots_concentration_QKM}. Likewise, it has been suggested that kernel concentration effects may strongly depend on the characteristics of the targeted dataset \cite{mirozewki2024_conf}, and various links have been established with respect to the depth and expressivity of the associated quantum circuits \cite{Peters2023generalization, Gil-Fuster_2024}. These topics deserve more attention and are currently among the most active areas of research in the QML field.

Quantum kernel methods can still offer better scaling than alternative QML variational approaches, in addition to the already mentioned guarantees in accessing true global optima \cite{schuld2021_are_kernel_methods}. In the longer term, if reliability of quantum hardware progresses enough (and while improving, this is still a big if), alternative implementations of quantum vector machines may become accessible, for which it is known that training time can grow linearly with the size of the data \cite{Rebentrost_2014}. 

In future work, we are committed to extend the current benchmark to include additional optimization strategies and benchmark datasets. In the first case, this includes comparing weighted versus unweighted alignment, multi-class and unbalanced problems, and exploring the suitability of additional quantum feature maps. We also anticipate the need to investigate the case for real-world and domain-specific applications that involve potentially more challenging data.

\section{Acknowledgment}
This study has been supported by project \text{RYC2022-038121-I}, funded by \lword{MCIN/AEI/10.13039/501100011033} and European Social Fund Plus (ESF+) and project \text{PID2023-147422OB-I00} funded by \lword{MCIU/AEI/10.13039/501100011033} and by the European FEDER program.

\bibliographystyle{ieeetr}
\bibliography{references}  






\end{document}